\documentclass[prb,twocolumn]{revtex4}

\usepackage{graphicx}
\usepackage{dcolumn}
\usepackage{bm}
\usepackage{bbm}

\begin{document}


\title{Effect of External Normal and Parallel Electric Fields \\ on $180^{\circ}$ Ferroelectric Domain Walls in PbTiO$_3$}

\author{Arzhang Angoshtari}
\author{Arash Yavari}
 \email{arash.yavari@ce.gatech.edu}
\affiliation{ School of Civil and Environmental Engineering, Georgia
Institute of Technology, Atlanta, GA 30332. }

\date{\today}



\begin{abstract}
We impose uniform electric fields both parallel and normal to
$180^{\circ}$ ferroelectric domain walls in PbTiO$_3$ and obtain the
equilibrium structures using the method of anharmonic lattice
statics. In addition to Ti-centered and Pb-centered perfect domain
walls, we also consider Ti-centered domain walls with oxygen
vacancies. We observe that electric field can increase the thickness
of the domain wall considerably. We also observe that increasing the
magnitude of electric field we reach a critical electric field
$E^c$; for $E > E^c$ there is no local equilibrium configuration.
Therefore, $E^c$ can be considered as an estimate of the threshold field $E_h$ for domain wall motion. Our
numerical results show that Oxygen vacancies decrease the value of
$E^c$. As the defective domain walls are thicker than perfect walls,
this result is in agreement with the recent experimental
observations and continuum calculations that show thicker domain
walls have lower threshold fields.
\end{abstract}


\maketitle

\section{Introduction}

Ferroelectric materials have been used in many important
applications such as high strain actuators, electro-optical systems,
non-volatile and high density memories, etc.
\cite{Scott2007,Kalinin2010}. The properties of domain walls in
ferroelectric materials including their structure, thickness, and
mobility are important parameters as they determine the
performance of devices that use these materials \cite{Jia2008}.

Theoretical calculations have predicted that ferroelectric domain walls
are atomically sharp and their thickness is about a few angstroms
\cite{MeyerVanderbilt2001,Padilla1996,YaOrBh2006b,AngYa2010}. However,
experimental measurements show the existence of domain walls with
thicknesses of a few micrometers \cite{Iwata2003,Lehnen2000}.
It has been observed that such broadening of domain walls is due to the
presence of extrinsic defects, charged walls, and surfaces
\cite{Choudhury2008}. Shilo \textit{et al.} \cite{Shilo2004} used
atomic force microscopy to measure the surface profile close to
emerging domain walls in PbTiO$_3$ and then fitted it to the
soliton-type solution of GLD theory. They measured wall widths of
$1.5nm$ and $4nm$ and observed a wide scatter in wall widths. They
suggested that the presence of point defects is responsible for such
wide variations. Lee \textit{et al.} \cite{Lee2005} proposed a continuum model
to investigate this proposal and reproduced the experimentally
observed range of wall widths with their model. They mentioned that
the interaction between the order parameter and point defects and
interaction of point defects with each other are two important
interactions that should be considered properly in such modelings.
Jia \textit{et al.} \cite{Jia2008} investigated the cation-oxygen
dipoles near $180^{\circ}$ domain walls in
PbZr$_{0.2}$Ti$_{0.8}$O$_{3}$ thin films. They measured the width
and dipole distortion across domain walls using the negative
spherical-aberration imaging technique in an aberration-corrected
transmission electron microscope and observed a large difference in
atomic details between charged and uncharged domain walls.

External electric field can cause the motion of ferroelectric domain
walls if the magnitude of the field reaches the threshold field
$E_h$ for wall motion, i.e., the field at which a domain wall
begins to move after overcoming the intrinsic Peierls friction of
the ferroelectric lattice \cite{Choudhury2008}. It was observed that
threshold fields that are predicted via thermodynamic calculations
are usually much greater than the experimental values. For example,
Bandyopadhyay and Ray \cite{BandyopadhyayRay2004} predicted an
upper limit for $E_h$ of LiNbO$_3$ to be $30000kV/cm$ but
experimental observations show that the threshold field for wall motion
can be less than $15kV/cm$. Choudhury \textit{et al.}
\cite{Choudhury2008} suggested that the reason for such large
differences between theoretical and experimental values of $E_h$ is
broadening of the domain walls. Using microscopic phase-field
modeling, they show that the threshold field for moving an
antiparallel ferroelectric domain wall dramatically drops by two or
three orders of magnitude if the wall was diffused by only about $1-2nm$. Su and Landis \cite{Landis2007} developed a continuum
thermodynamics framework to model the evolution of ferroelectric
domain structures and investigated the fields near $90^{\circ}$ and
$180^{\circ}$ domain walls and the electromechanical pining strength
of an array of line charges on these domain walls.

In this work, we investigate the effect of external electric field
$(E)$ on the perfect and defective $180^{\circ}$ domain walls in
PbTiO$_3$ using the method of anharmonic lattice statics. We
consider both Pb-centered and Ti-centered perfect domain walls and
also defective domain walls with oxygen vacancies. In agreement with
experimental results, our calculations show that such defective domain
walls are thicker than perfect walls \cite{AngYavari2010}. By
increasing $E$ we reach a critical value $E^c$ such
that for $E>E^c$ the lattice statics iterations do not converge.
Therefore, this critical value can be considered as a lower bound
for the threshold field for wall motion.

This paper is organized as follows. In \S\ref{sec:Ferro}, we explain
the geometry of the perfect and defective domain walls that we use
throughout this work. In \S\ref{sec:Calculation}, we describe the
method of analysis used in our calculations. Our
numerical results are presented in \S\ref{sec:Results}. The paper ends with
some concluding remarks in \S\ref{sec:Concluding}.

\begin{figure*}[t]
\centerline{\hbox{\includegraphics[scale=0.8,angle=0]{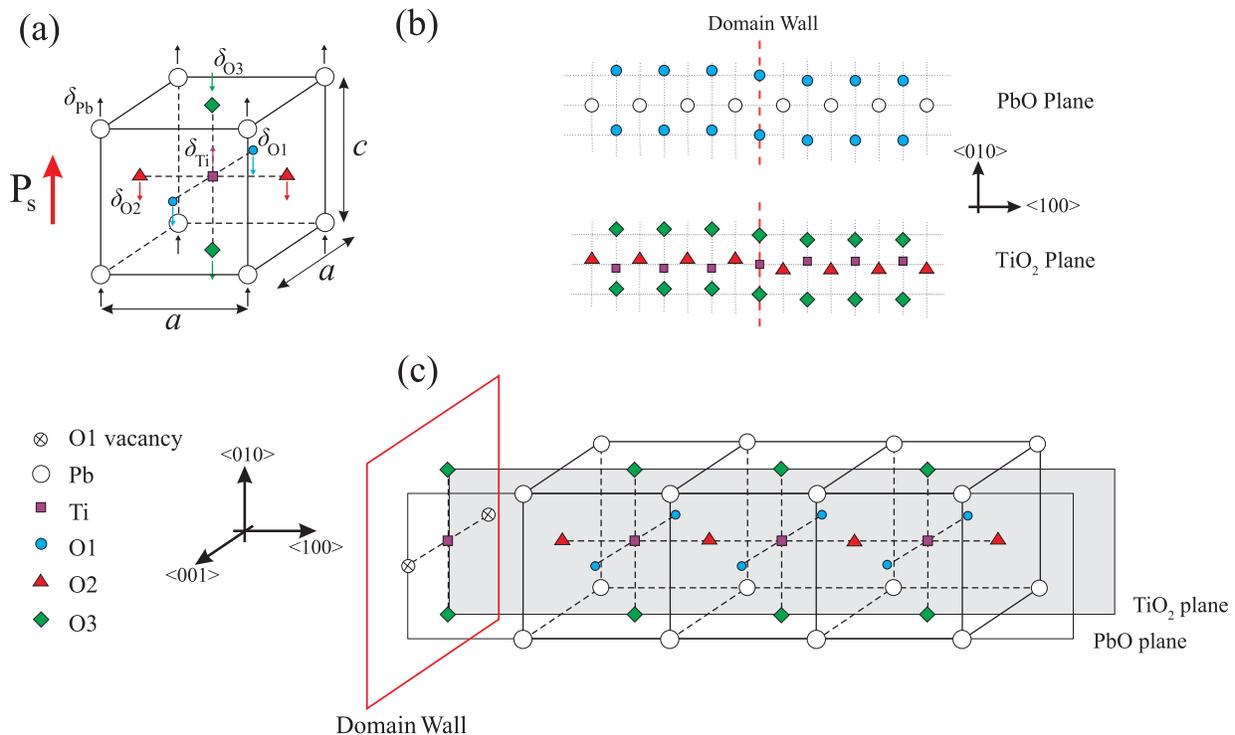}
\vspace*{-0.4in} }}
 \caption{ (a) The relaxed configuration of the
unit cell of PbTiO$_3$. $a$ and $c$ are the tetragonal lattice
parameters.  Note that O1, O2, and O3 refer to oxygen atoms located
on $(001)$, $(100)$, and $(010)$-planes, respectively. $\delta$
denotes the y-displacements of the atoms from their centerosymmetric
positions and arrows near each atom denote the direction of these
displacements. (b) The geometry of a perfect Ti-centered
$180^{\circ}$ domain wall. (c) The geometry of an O1-defective
$180^{\circ}$ domain wall. Note that Pb-centered domain walls with
oxygen vacancies are not stable.} \label{Geom}
\end{figure*}

\section{Ferroelectric Domain Walls} \label{sec:Ferro}

Due to the relative displacements between the center of the
positive and negative charges, each unit cell of a ferroelectric
crystal has a net polarization below its Curie temperature. Fig.\ref{Geom}(a) shows the relaxed unit cell of tetragonal PbTiO$_3$.
In this work, we consider $180^{\circ}$ domain walls in PbTiO$_3$
parallel to a $(100)$-plane. These domain walls are two dimensional
defects across which the direction of the polarization vector switches. There are two types of perfect $180^{\circ}$ domain walls in
PbTiO$_3$: Pb-centered and Ti-centered domain walls. Fig.\ref{Geom}(b) shows the geometry of a Ti-centered domain wall.

In addition to perfect domain walls, we also consider $180^{\circ}$
domain walls with oxygen vacancies. It is known that oxygen
vacancies tend to move toward domain walls and pin them
\cite{HeVanderbilt2001,Calleja2003,Salje2010}. Therefore, we study
domain walls with oxygen vacancies sitting on them. In
order to be able to obtain a solution, we need to consider
periodically arranged vacancies on the domain walls. Although in
reality oxygen vacancies have lower densities, our results with the
current assumption can still provide important insights on the
effect of oxygen vacancies on $180^{\circ}$ domain walls. Depending on
which oxygen in the PbTiO$_3$ unit cell sits on the domain wall,
there would be three types of defective domain walls: (i)
O2-defective, (ii) O1-defective, and (iii) O3-defective. Fig.\ref{Geom}(c) shows an O1-defective domain wall. Note that O1- and
O3-defective domain walls are Ti-centered while O2-defective domain
wall is Pb-centered. It has been observed that O2-defective domain walls
are not stable \cite{AngYavari2010,HeVanderbilt2001}, i.e., the
lattice statics iterations do not converge. Thus, we consider O1-
and O3-defective domain walls in the following.

Let x, y, and z denote coordinates along the $\langle100\rangle$,
$\langle010\rangle$, and $\langle001\rangle$-directions,
respectively. We assume a 1D symmetry reduction, which means that
all the atoms with the same x-coordinates have the same
displacements. Therefore, we partition the 3D lattice $\mathcal{L}$
as $\mathcal{L}=\bigsqcup_{I}^{}\bigsqcup_{\alpha\in
\mathbbm{Z}}\mathcal{L}_{I\alpha}$, where $\mathcal{L}_{I\alpha}$
and $\mathbbm{Z}$ are 2D equivalence classes parallel to the
$(100)$ plane and the set of integers, respectively. $j=J\beta$
is the atom in the $\beta$\emph{th} equivalence class
of the $J$\emph{th} sublattice. See
\cite{YaOrBh2006a,YavariAngoshtari2010} for more details on the
symmetry reduction.

\section{Method of Calculation}\label{sec:Calculation}

We apply a uniform electric field on $180^{\circ}$ domain walls and
obtain the equilibrium structure using the method of anharmonic
lattice statics \cite{YaOrBh2006a}. We use a shell potential for
PbTiO$_3$ \citep{Asthagiri2006} for modeling the atomic
interactions. Each ion is represented by a core and a massless shell
in this potential. Let $\mathcal{L}$ denote the collection of cores
and shells, $i\in\mathcal{L}$ denotes a core or a shell in
$\mathcal{L}$, and $\left\{\mathbf{x}^i \right\}_{i\in \mathcal{L}}$
represents the current position of cores and shells. In this shell
potential, three different energies are assumed to exist due to the
interactions of cores and shells: $\mathcal{E}_{\textrm{short}}$,
$\mathcal{E}_{\textrm{long}}$, and
$\mathcal{E}_{\textrm{core-shell}}$.
$\mathcal{E}_{\textrm{short}}\left(\left\{\mathbf{x}^i
\right\}_{i\in \mathcal{L}}\right)$ denotes the energy of short range
interactions, which are assumed to be only between Pb-O, Ti-O, and
O-O shells. The short range interactions are described by the
Rydberg potential of the form $(A+Br)\exp(-r/C)$, where A, B, and C
are potential parameters and $r$ is the distance between interacting
elements. $\mathcal{E}_{\textrm{long}}\left(\left\{\mathbf{x}^i
\right\}_{i\in \mathcal{L}} \right)$ denotes the Coulombic
interactions between the core and shell of each ion with the cores
and shells of all the other ions. For calculating
the classical Coulombic energy and force, we use the damped Wolf
method \cite{Wolf99}. Finally,
$\mathcal{E}_{\textrm{core-shell}}\left(\left\{\mathbf{x}^i
\right\}_{i\in \mathcal{L}}\right)$ represents the interaction of
core and shell of an atom and is assumed to be an anharmonic spring
of the form $(1/2)k_2 r^2 +(1/24)k_4 r^4$, where $k_2$ and $k_4$ are
constants. The total static energy is written as
\begin{eqnarray}
    && \mathcal{E}\left(\left\{\mathbf{x}^i \right\}_{i\in \mathcal{L}}
    \right) = \mathcal{E}_{\textrm{short}}\left(\left\{\mathbf{x}^i \right\}_{i\in \mathcal{L}}
    \right)
    + \mathcal{E}_{\textrm{long}}\left(\left\{\mathbf{x}^i \right\}_{i\in
    \mathcal{L}} \right) \nonumber \\
    && ~~~~~~~~~~~~~~~~~~~+\mathcal{E}_{\textrm{core-shell}}\left(\left\{\mathbf{x}^i \right\}_{i\in \mathcal{L}}
    \right).
\end{eqnarray}
Note that all the calculations are done for absolute zero temperature.
At this temperature PbTiO$_3$ has a tetragonal unit cell with
lattice parameters $a=3.843~{\AA}$ and $c=1.08a$
\cite{Asthagiri2006}.

Assume that a uniform electric field $\mathbf{E}=(E_{x},E_{y},E_{z})$ is
applied to a collection of atoms. Then for the relaxed configuration
$\mathcal{B}=\left\{\mathbf{x}^i
\right\}_{i\in\mathcal{L}}\subset\mathbbm{R}^3$, we have
\begin{equation}\label{equilibrium}
    \frac{\partial \mathcal{E}}{\partial \mathbf{x}^i}+
    q_{i}\mathbf{E}=\mathbf{0}~~~~~~~\forall~i\in\mathcal{L},
\end{equation}
where $q_i$ denotes the charge of the i\emph{th} particle (core or shell). To obtain the solution of the above problem, we utilize the
Newton method. Having a configuration $\mathcal{B}^{k}$ the next configuration $\mathcal{B}^{k+1}$ is
calculated from the current configuration $\mathcal{B}^{k}$ as:
$\mathcal{B}^{k+1}=\mathcal{B}^{k}+\tilde{\boldsymbol{\delta}}^k$,
where
\begin{equation}
    \tilde{\delta}^k=-\mathbf{H}^{-1}\left(\mathcal{B}^{k}\right)\cdot\boldsymbol{\nabla}\mathcal{E}\left(\mathcal{B}^{k}\right),
\end{equation}
with $\mathbf{H}$ denoting the Hessian matrix. The calculation of the Hessian becomes inefficient as the size of
the problem increases and hence we use the quasi-Newton method.
This method uses the Broyden-Fletcher-Goldfarb-Shanno (BFGS)
algorithm to approximate the inverse of the Hessian
\cite{PressTVF1989} instead of the direct calculation of the Hessian
at each iteration. We start from a positive-definite matrix and use
the BFGS algorithm to update the Hessian at each iteration as
follows:
\begin{eqnarray}\label{BFGS}
    && \mathbf{C}^{i+1} = \mathbf{C}^i + \frac{\tilde{\boldsymbol{\delta}}^k\otimes\tilde{\boldsymbol{\delta}}^k}{
    (\tilde{\boldsymbol{\delta}}^k)^{\textsf{T}}\cdot\mathbf{\Delta}}
    -\frac{\left(\mathbf{C}^{i}\cdot\mathbf{\Delta}\right)\otimes\left(\mathbf{C}^{i}\cdot\mathbf{\Delta}\right)}
          {\mathbf{\Delta}^{\textsf{T}}\cdot\mathbf{C}^{i}\cdot\mathbf{\Delta}} \nonumber\\
    && ~~~~~~~~+ \left(\mathbf{\Delta}^{\textsf{T}}\cdot\mathbf{C}^{i}\cdot\mathbf{\Delta}\right)\mathbf{u}\otimes\mathbf{u},
\end{eqnarray}
where $\mathbf{C}^{i}=\left(\mathbf{H}^i\right)^{-1}$,
$\mathbf{\Delta}=\boldsymbol{\nabla}\mathcal{E}^{i+1}-\boldsymbol{\nabla}\mathcal{E}^{i}$, and
\begin{eqnarray}
  \mathbf{u}=\frac{\tilde{\boldsymbol{\delta}}^k}{(\tilde{\boldsymbol{\delta}}^k)^{\textsf{T}}\cdot\mathbf{\Delta}}-
  \frac{\mathbf{C}^{i}\cdot\mathbf{\Delta}}{\mathbf{\Delta}^{\textsf{T}}\cdot\mathbf{C}^{i}\cdot\mathbf{\Delta}}.
\end{eqnarray}
Calculating $\mathbf{C}^{i+1}$, one then should use
$\mathbf{C}^{i+1}$ instead of $\mathbf{H}^{-1}$ to update the
current configuration. If $\mathbf{C}^{i+1}$ is a poor approximation, then one may need to
perform a linear search to refine $\mathcal{B}^{k+1}$ before
starting the next iteration \cite{PressTVF1989}.
\begin{figure*}[t]
\centerline{\hbox{\includegraphics[scale=0.8,angle=0]{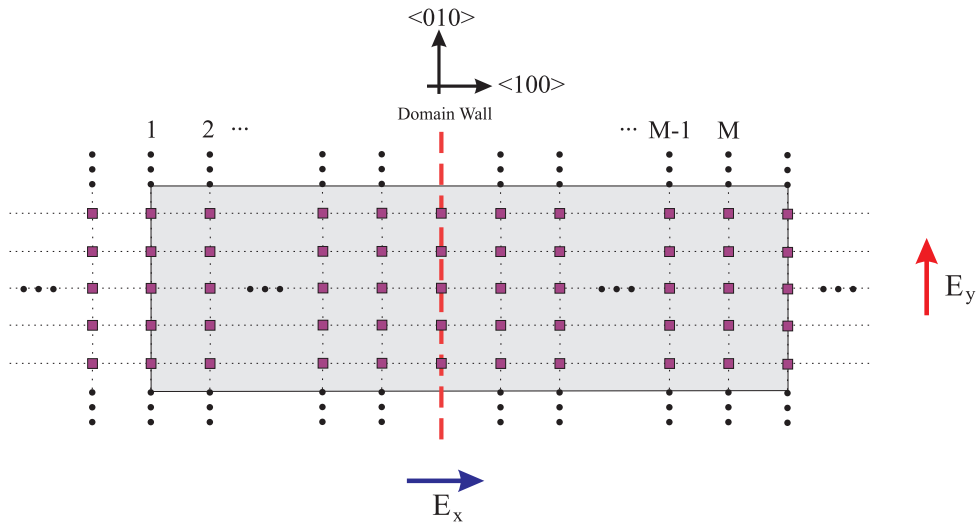}
\vspace*{-0.2in}
                 }}
\caption{Ti-cores under external electric field in a Ti-centered
$180^{\circ}$ domain wall. $E_x$ and $E_y$ are the normal and
parallel electric fields, respectively. The shaded region denotes
the region that is relaxed in each step. Note that $M$ is the size
of the computational box (CB) normal to the domain wall. We consider
different CBs with one, four, and sixteen unit cells in
the domain wall plane. \label{CB}}
\end{figure*}

\begin{figure*}[t]
\centerline{\hbox{\includegraphics[scale=.8,angle=0]{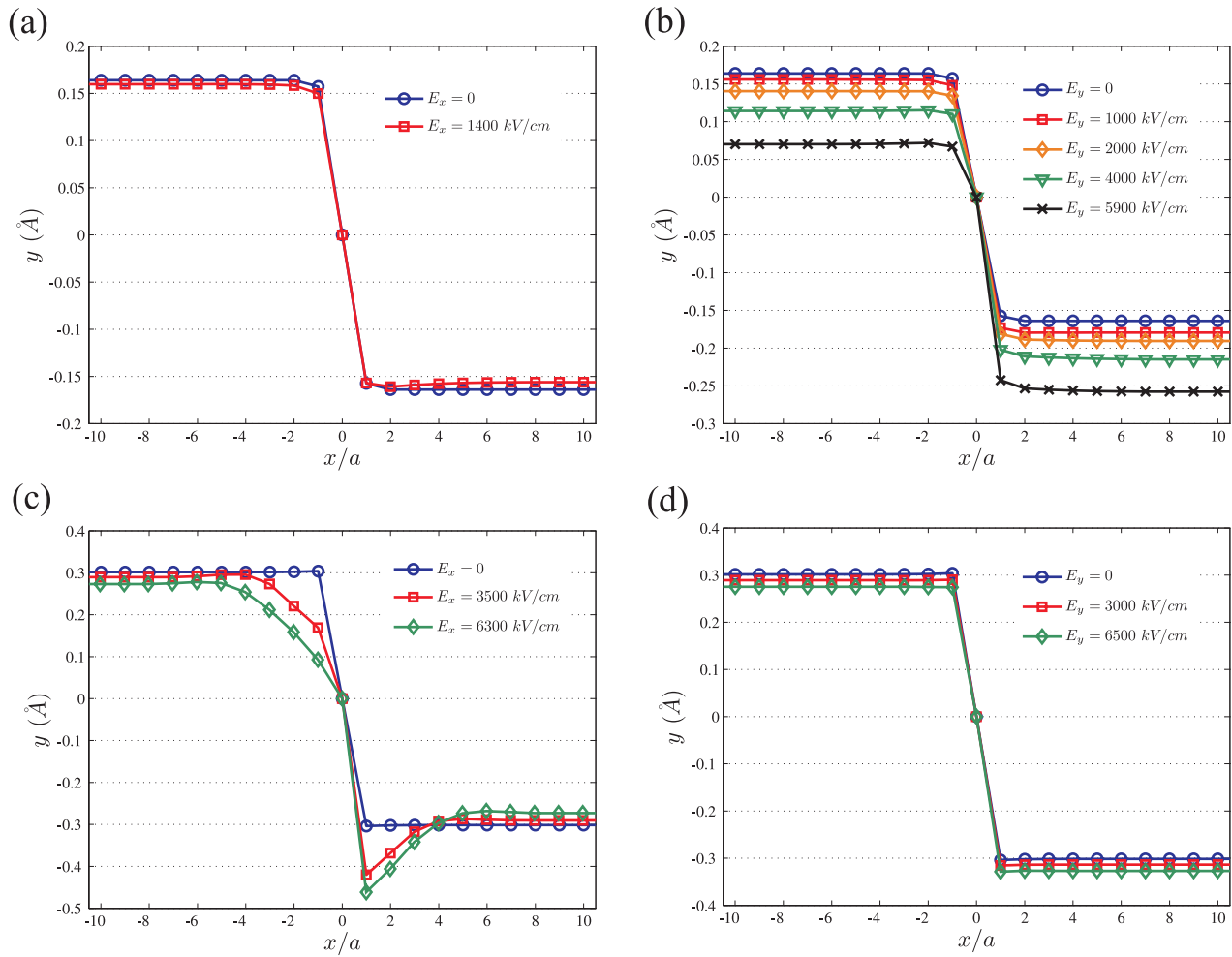}
\vspace*{-0.2in}
                 }}
\caption{The y-coordinates of cores under external electric field:
Ti cores in a perfect Ti-centered domain wall under (a) $E_x$ and
(b) $E_y$; Pb cores in a perfect Pb-centered domain wall under (c)
$E_x$ and (d) $E_y$. \label{Perfect}}
\end{figure*}

\begin{figure*}[t]
\centerline{\hbox{\includegraphics[scale=.85,angle=0]{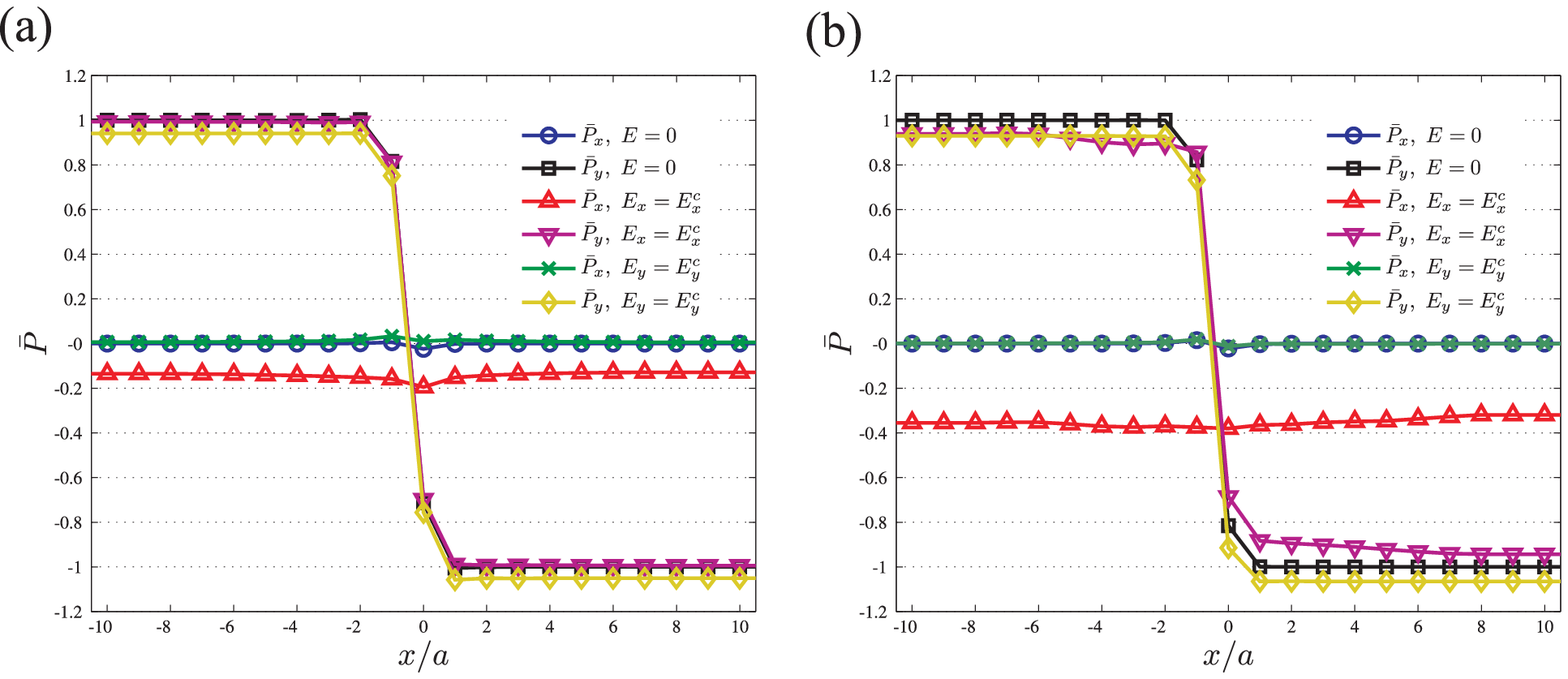}
\vspace*{-0.2in}
                 }}
\caption{The polarization profiles
$\overline{\mathbf{P}}=(\bar{P}_x,\bar{P}_y)$ of domain walls under
zero, normal critical field ($E^{c}_{x}$), and parallel critical
field ($E^{c}_{y}$) for (a) Ti-centered, and (b) Pb-centered domain
walls. \label{PolarPerfect}}
\end{figure*}

In the presence of oxygen vacancies on the domain wall, one needs to
consider charge redistribution between some ions. To model an oxygen
vacancy using a shell potential, we remove the core and shell of the
oxygen atom and because we assume a charge neutral oxygen vacancy,
there will be a charge redistribution in the neighboring shells
\cite{AngYavari2010}. It is known that charge redistribution is
highly localized and hence in our calculations we equally
distribute the charge $\Delta Q=Q_s+Q_c$, where $Q_s$ and $Q_c$ are
oxygen shell and core charges, between the (fourteen) first nearest
neighbors of each oxygen vacancy.
\begin{figure*}[t]
\centerline{\hbox{\includegraphics[scale=.8,angle=0]{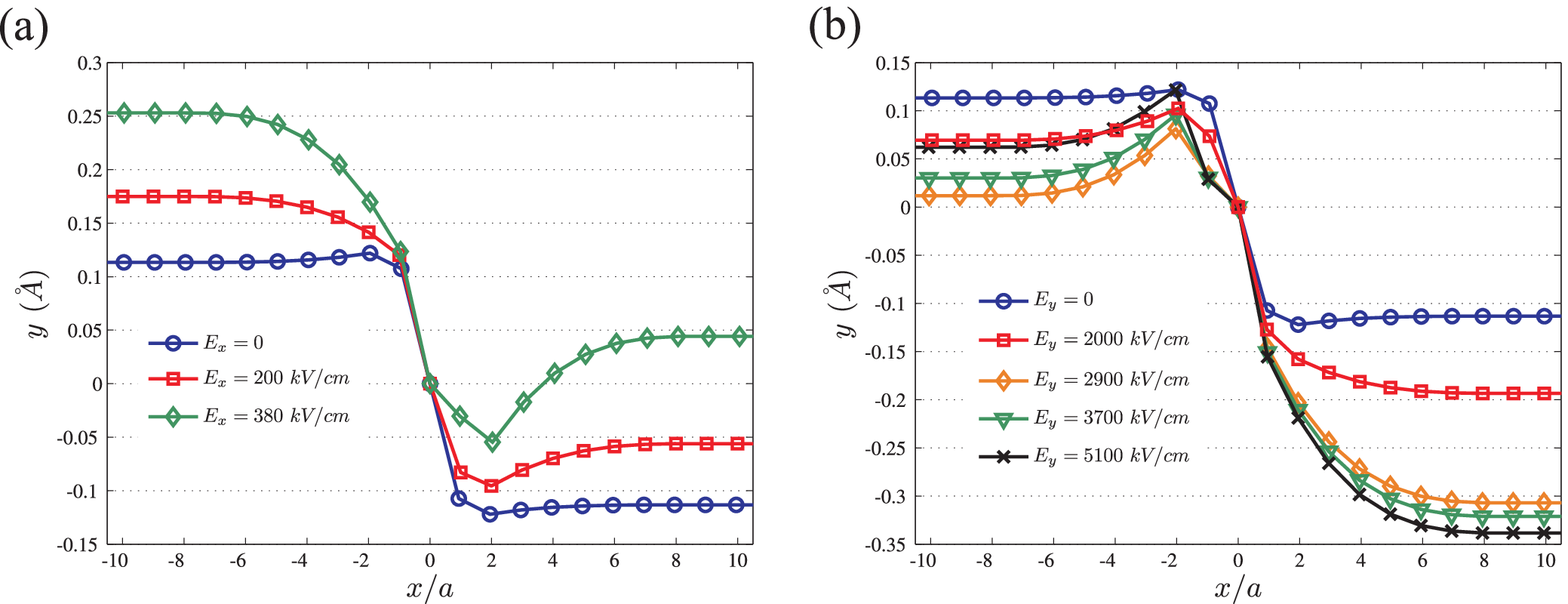}
\vspace*{-0.2in}
                 }}
\caption{The y-coordinates of Ti cores in an O1-defective domain wall
under (a) normal ($E_x$), and  (b) parallel ($E_y$) external electric
fields. \label{O1}}
\end{figure*}

\begin{figure}[t]
\begin{center}
\includegraphics[scale=0.65,angle=0]{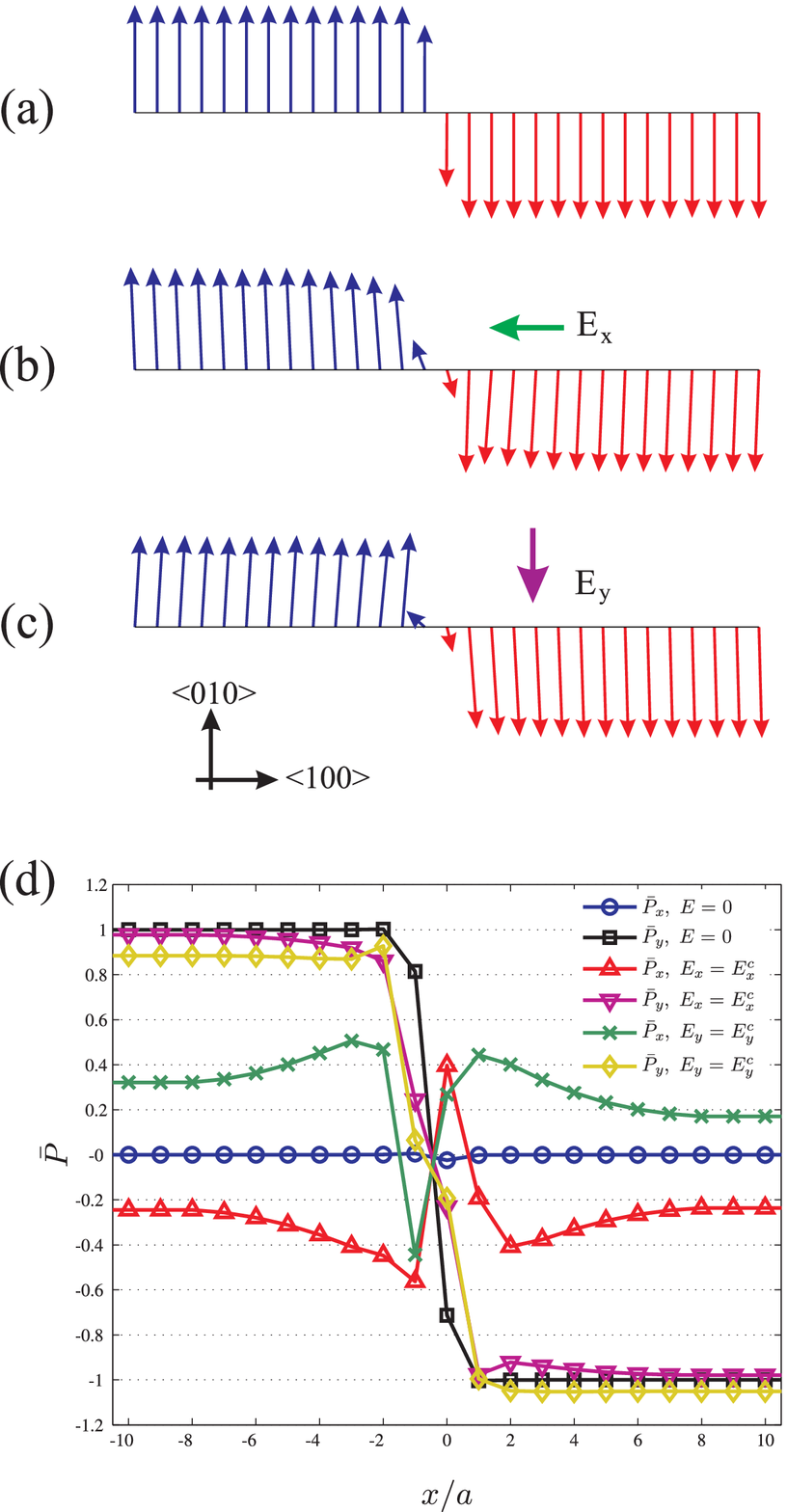}
\end{center}\vspace*{-0.2in}
\caption{The polarization profiles of O1-defective domain walls
under (a) zero, (b) normal critical field ($E^{c}_{x}$), and (c)
parallel critical field ($E^{c}_{y}$). (d) Components of the
polarization vector $\overline{\mathbf{P}}=(\bar{P}_x,\bar{P}_y)$ of
O1-defective domain walls under zero, $E^{c}_{x}$, and $E^{c}_{y}$.}
\label{PolarO1}
\end{figure}

To obtain the equilibrium configuration under an external electric
field we need to start from an appropriate initial configuration.
This initial configuration for perfect and defective domain walls is
the equilibrium configuration of these domain walls under zero
electric field (see \cite{AngYavari2010,YaOrBh2006b} for discussions
on how to calculate these configurations). As we mentioned earlier,
we assume a 1D symmetry reduction for the lattice and hence as is
shown in Fig.\ref{CB}, our computational box (CB) consists of a row
of unit cells perpendicular to the domain wall. In this figure, the
shaded region is the computational box. Note that because in general
there is no symmetry in the problem, we need to relax all the atoms
inside the CB. For removing the rigid body translation freedom of
the atoms, one should fix the core of an atom and relax the other
atoms. We fix Pb-core (Ti-core) of an atom located on the domain
wall in Pb-centered (Ti-centered) domain walls. Thus, if there are
$M$ unit cells in the CB, we would have $30M-3$ variables in our
calculations. We should mention that to investigate the effect of the size of CB
in the domain wall plane, we consider CBs with one, four,
and sixteen unit cells in the domain wall plane and therefore the
number of the unit cells in CB in each case is $M$, $4M$, and $16M$,
respectively. We observe that the final relaxed structure does not
depend on the size of CB in the domain wall plane. This suggests
that the symmetry reduction that we use in our calculations is a
reasonable assumption for this problem.

Note that we consider a finite number of unit cells in the CB
and do not assume any periodicity condition in our calculations.
This means that we need to impose some proper boundary conditions to take into account the
effect of the atoms located outside of CB. To this end, we
rigidly move the unit cells outside of CB with displacements equal to
those of the first or last unit cell of the CB (the unit
cell on the boundary of the CB that is closer to the unit cell
outside of the CB). This is a natural boundary condition as we expect
the bulk configuration far from the domain wall.

In our calculations we use $M=20$ as larger values for $M$ do
not affect the results. Imposing an external electric field should
be done step by step, i.e., one first needs to obtain the
configuration for $\mathbf{E}=\Delta \mathbf{E}_{1}$ from the
initial configuration and then use this configuration to obtain the
equilibrium configuration for $\mathbf{E}=\Delta
\mathbf{E}_{1}+\Delta \mathbf{E}_{2}$ and so on. We use the average
step size of $20kV/cm$ for electric field. Using this step size and
force tolerance of $0.005eV{\AA}^{-1}$, our solutions converge
after about $30$ to $40$ iterations.

\section{Numerical Results }\label{sec:Results}

In this section we present our numerical results for perfect and
defective domain walls. Note that as the coordinates of cores and
shells are close to each other, we only report the results for
cores. Also as we mentioned earlier, x, y, and z are coordinates
along the $\langle100\rangle$, $\langle010\rangle$, and
$\langle001\rangle$-directions, respectively.

\textbf{Perfect domain walls:} We plot the y-coordinates of Ti-cores
under external electric field normal to the Ti-centered domain wall,
$E_x$, in Fig.\ref{Perfect}(a). As expected, we see that
increasing the electric field, the atomic structure
loses its symmetry. We observe that there exists an upper bound for $E_x$, i.e., there exists a critical electric field
$E^{c}_{x}$ such that for $E_{x}>E^{c}_{x}$ there is no local
equilibrium structure. The critical value of the normal electric
field is about $E^{c}_{x}=1400kV/cm$. The thickness of the domain
wall slightly increases as the normal electric
field increases. Note that domain wall thickness cannot be defined
uniquely very much like boundary layer thickness in fluid mechanics.
Here, domain wall thickness is by definition the region that is
affected by the domain wall, i.e. those layers of atoms that are
distorted. One can use definitions like the $99\%$-thickness in
fluid mechanics and define the domain wall thickness as the length
of the region that has $99\%$ of the far field rigid translation
displacement. What is important here is that no matter what
definition is chosen, domain wall ``thickness" increases as the
normal electric field increases. For a Ti-centered domain wall, the domain
wall thickness increases from $3$ atomic spacings ($1nm$) to
about $5$ atomic spacings ($1.5nm$) for $E_{x}=E^{c}_{x}$.

Fig.\ref{Perfect}(b) depicts the y-coordinates of Ti-cores under an
external electric field $E_y$ parallel to a Ti-centered domain wall.
It is observed that such electric fields do not alter the domain
wall thickness. Note that similar to the atomic structure for normal
fields, the atomic structure under parallel fields loses its
symmetry as well. The critical value of the parallel electric field is about
$E^{c}_{y}=5900kV/cm$, which is $4$ times larger than that of the
normal electric field.

Fig.\ref{Perfect}(c) shows the y-coordinates of Pb-cores of a
Pb-centered domain wall under normal electric field $E_x$. We
observe that the critical electric field is about
$E^{c}_{x}=6300kV/cm$, which is about $4.5$ times greater than the
critical normal electric field of Ti-centered walls. Also it is observed that
domain wall thickness increases to about $11$ atomic spacings ($4nm$) under critical normal electric field. The y-coordinates of Pb-cores of
a Pb-centered domain wall under parallel electric field $E_y$ are
shown in Fig.\ref{Perfect}(d). Similar to perfect Ti-centered domain
walls, we observe that parallel electric fields do not affect the
domain wall thickness. The critical parallel electric field is about
$E^{c}_{y}=6500kV/cm$. For Pb-centered domain walls we see that
unlike Ti-centered domain walls, the critical normal electric field
is close to the critical parallel electric field.

Fig.\ref{PolarPerfect} depicts the polarization profiles normal and
parallel to the domain walls. For calculation of the cell-by-cell
polarization, we follow Meyer and Vanderbilt
\cite{MeyerVanderbilt2001}. We plot
$\overline{\mathbf{P}}=(\bar{P}_{x},\bar{P}_{y})=
\mathbf{P}/|\mathbf{P}_{b}|$, where $\mathbf{P}$ is the polarization
and $|\mathbf{P}_{b}|=80.1\mu Ccm^{-2}$ is the norm of the bulk
polarization \cite{AngYa2010_4}. Fig.\ref{PolarPerfect}(a) shows
$\bar{P}_{x}$ and $\bar{P}_{y}$ for a Ti-centered domain walls under
zero and critical electric fields. In agreement with Lee \textit{et
al.} \cite{Lee2009} and Angoshtari and Yavari \cite{AngYa2010}, it
is observed that $(100)$ Ti-centered domain walls have
a mixed Ising-N\'{e}el character, i.e., polarization rotates normal
to the $(100)$-plane near the domain wall. For $E=0$, the maximum
normal component of the polarization is about $2\%$ of the bulk
polarization. For $E=E^{c}_{x}$, as can be expected, normal electric field
causes the positive and negative charges to have normal displacements
that create a polarization in the x-direction. This normal
component of the polarization ($\bar{P}_x$) reaches to about
$13.5\%$ of the bulk polarization at $E^{c}_x$, but we observe that
normal electric field $E^{c}_x$ does not have a remarkable effect on the
parallel component of polarization, $\bar{P}_{y}$. On the other
hand, we observe that under $E=E^{c}_{y}$, $\bar{P}_{x}$ does not
change considerably but $\bar{P}_y$ has an unsymmetric profile
with the maximum value of about $105\%$ of the bulk polarization.

Fig.\ref{PolarPerfect}(b) presents similar results for Pb-centered
domain walls. Similar to Ti-centered domain walls, we observe that
Pb-centered domain walls have a mixed Ising-N\'{e}el character
\cite{Lee2009,AngYa2010} with $\bar{P}_x$ about $2\%$ of the bulk
polarization for zero electric field. For $E=E^{c}_{x}$,
$\bar{P}_{x}$ reaches to about $38\%$ of the bulk polarization. Also
we observe that $E^{c}_{x}$ has more impact on $\bar{P}_{y}$
compared to Ti-centered walls. Finally, it is observed that
similar to Ti-centered domain walls, $E^{c}_{y}$ does not have a
significant effect on $\bar{P}_x$ but makes $\bar{P}_y$ unsymmetric
with maximum value of about $107\%$ of the bulk polarization.

\textbf{Defective domain walls:} In this part we report the
structure of defective domain walls under normal and parallel
external electric fields. Because the results for O1- and O3-defective
domain walls are similar, we only present the results for
O1-defective walls, which are Ti-centered. Note that as
we mentioned earlier, O2-defective domain walls, which are Pb-centered, are not stable. Our calculations show that they are not stable even under external electric fields. We had earlier shown that they are not stable under strain as well \cite{AngYavari2010}.

Fig.\ref{O1}(a) depicts the y-coordinates of Ti-cores in an
O1-defective domain wall under normal external electric field $E_x$.
The critical normal field is about $E^{c}_{x}=380kV/cm$. It is
observed that domain wall thickness increases up to about $16$
atomic spacings ($6nm$) under the critical normal electric
field. Comparing O1-defective atomic structure with the structure of
perfect Ti-centered domain wall under normal field
(Fig.\ref{Perfect}(a)), we observe that oxygen vacancies increase
the thickness of the domain wall considerably. Also it is observed
that critical normal electric field of defective domain walls is smaller than
that of perfect Ti-centered wall. Fig.\ref{O1}(b) shows the
y-coordinates of Ti-cores in an O1-defective domain walls under
parallel electric field $E_y$. The value of the critical field is
about $E^{c}_{y}=5100kV/cm$. Here we observe a major difference
between the atomic structures of perfect and defective domain walls;
unlike perfect domain walls, parallel electric fields increase the
thickness of defective domain walls up to about $13$ atomic spacings
($5nm$) under the critical parallel electric field. Also
similar to normal electric fields, we observe that critical electric
field of defective domain walls is smaller than that of perfect
domain walls.

Defective domain walls are thicker than perfect domain walls. The
observation that the defective domain walls have smaller critical
electric fields is in agreement with the experimental observations of
Choudhury \textit{et al.} \cite{Choudhury2008}. They observed that
the threshold field for domain wall motion exponentially decreases
as the domain wall width increases.

Fig.\ref{PolarO1} shows the polarization profiles for O1-defective
domain walls. It is observed that similar to perfect domain walls,
defective domain walls have an Ising-N\'{e}el character with
$\bar{P}_{x}$ of about $2.5\%$ of the bulk polarization for
zero electrical field. For $E=E^{c}_x$, $\bar{P}_x$ reaches to about
$55\%$ of the bulk polarization, which is greater than the
corresponding values for perfect domain walls, and $\bar{P}_y$ shows
more Ising-type character. As we mentioned earlier, for $E=E^{c}_y$
we observe a difference between perfect and defective domain
walls; unlike perfect walls, parallel electric fields have
considerable effects on $\bar{P}_x$: it reaches to about $55\%$ of
the bulk polarization under $E^{c}_y$. Similar to perfect domain walls,
$\bar{P}_y$ has an unsymmetric distribution and reaches to about
$106\%$ of the bulk polarization.

\section{Concluding Remarks}\label{sec:Concluding}

In this work we obtained the atomic structure of perfect and
defective $180^{\circ}$ domain walls in PbTiO$_3$ under both parallel and normal external
electric fields using the method of anharmonic lattice statics. We observe that
electric field can increase the thickness of a domain wall
considerably (up to $5$ times thicker than domain walls under no
external electric field). This can be one reason for the wide
scatter of the domain wall thicknesses observed in experimental
measurements. In agreement with previous works
\cite{AngYavari2010,Shilo2004}, we observe that oxygen vacancies can
increase the thickness of the domain walls. We also observe that by
increasing the external electric field we reach a
critical electric field $E^c$. For $E>E^c$ there is no local
equilibrium configuration and hence $E^c$ can be considered as an estimate of the threshold field for the domain wall motion. We observe
that defective domain walls, which are thicker than perfect domain
walls, have smaller critical fields. This is in agreement with the
experimental observations that show the threshold field decreases as
the domain wall thickness increases \cite{Choudhury2008}.

In practice, it has been observed that domain walls move or break down
under electric fields in the order of a few $kV/cm$
\cite{Choudhury2008,Roy2010}, which are considerably smaller than
the high-fields that we consider here. We do not consider break down
of the domain walls in our model. Also as was mentioned earlier, the
high density of oxygen vacancies that we assume is unrealistic.
In practice, steps and other complex defects on domain walls can
increase the thickness of the domain walls considerably
\cite{Iwata2003,Lehnen2000,AngYa2010_4}. Therefore, as the threshold
fields for domain walls decrease exponentially with the increase of
the domain wall width \cite{Choudhury2008}, one can obtain better
estimates for the critical electric fields with more realistic
models for defects on domain walls. Also as suggested by Roy
\textit{et al.} \cite{Roy2010} electric fields change the potential
parameters. In this paper our aim is to demonstrate that even with our simple model,
one can show that the threshold field has an inverse relation with
the domain wall thickness.

\acknowledgments We benefited from a discussion with C.M. Landis. We thank an anonymous reviewer whose comments helped us improve the paper.

\end{document}